\documentclass[12pt,a4paper]{article}
\pdfoutput=1
\usepackage[utf8]{inputenc}
\usepackage[english]{babel}
\usepackage{amsmath}
\usepackage{amsfonts}
\usepackage{amssymb}
\usepackage[margin=1in]{geometry}
\usepackage{mathptmx}
\usepackage{graphicx}
\usepackage[usenames,dvipsnames]{xcolor}
\usepackage{cite}
\usepackage{authblk}
\usepackage{soul}

\providecommand{\keywords}[1]{\textbf{\textit{Keywords:}} #1}
\title{Phase behaviour of DNA in\\ presence of DNA-binding proteins}

\author[1,2]{Guillaume Le Treut}
\author[2]{Fran\c{c}ois K\'ep\`es}
\author[1]{Henri Orland}
\affil[1]{Institut de Physique Th\'eorique, Universit\'e Paris Saclay, CEA, CNRS, F-91191 Gif-sur-Yvette, France}
\affil[2]{institute of Systems and Synthetic Biology, University of Evry-Val-d'Essonne,  CNRS, Genopole Campus 1, B\^at. 6, F-91030 Evry Cedex, France}

\date{}
\begin{document}

\maketitle

\begin{abstract}
To characterize the thermodynamical equilibrium of DNA chains interacting with a solution of non-specific binding proteins, a Flory-Huggins free energy model was implemented. We explored the dependence on DNA and protein concentrations of the DNA collapse. For physiologically relevant values of the DNA-protein affinity, this collapse gives rise to a biphasic regime with a dense and a dilute phase; the corresponding phase diagram was computed. Using an approach based on Hamiltonian paths, we show that the dense phase has either a molten globule or a crystalline structure, depending on the DNA bending rigidity, which is influenced by the ionic strength. These results are valid at the thermodynamical equilibrium and should therefore be consistent with many biological processes, whose characteristic timescales range typically from 1 ms to 10 s. Our model may thus be applied to biological phenomena that involve DNA-binding proteins, such as DNA condensation with crystalline order, which occurs in some bacteria to protect their chromosome from detrimental factors; or transcription initiation, which occurs in clusters called transcription factories that are reminiscent of the dense phase characterized in this study.
\end{abstract}


\newpage
\section{Introduction}

\begin{figure}[ht!]
\centering
\includegraphics[width=0.75 \linewidth]{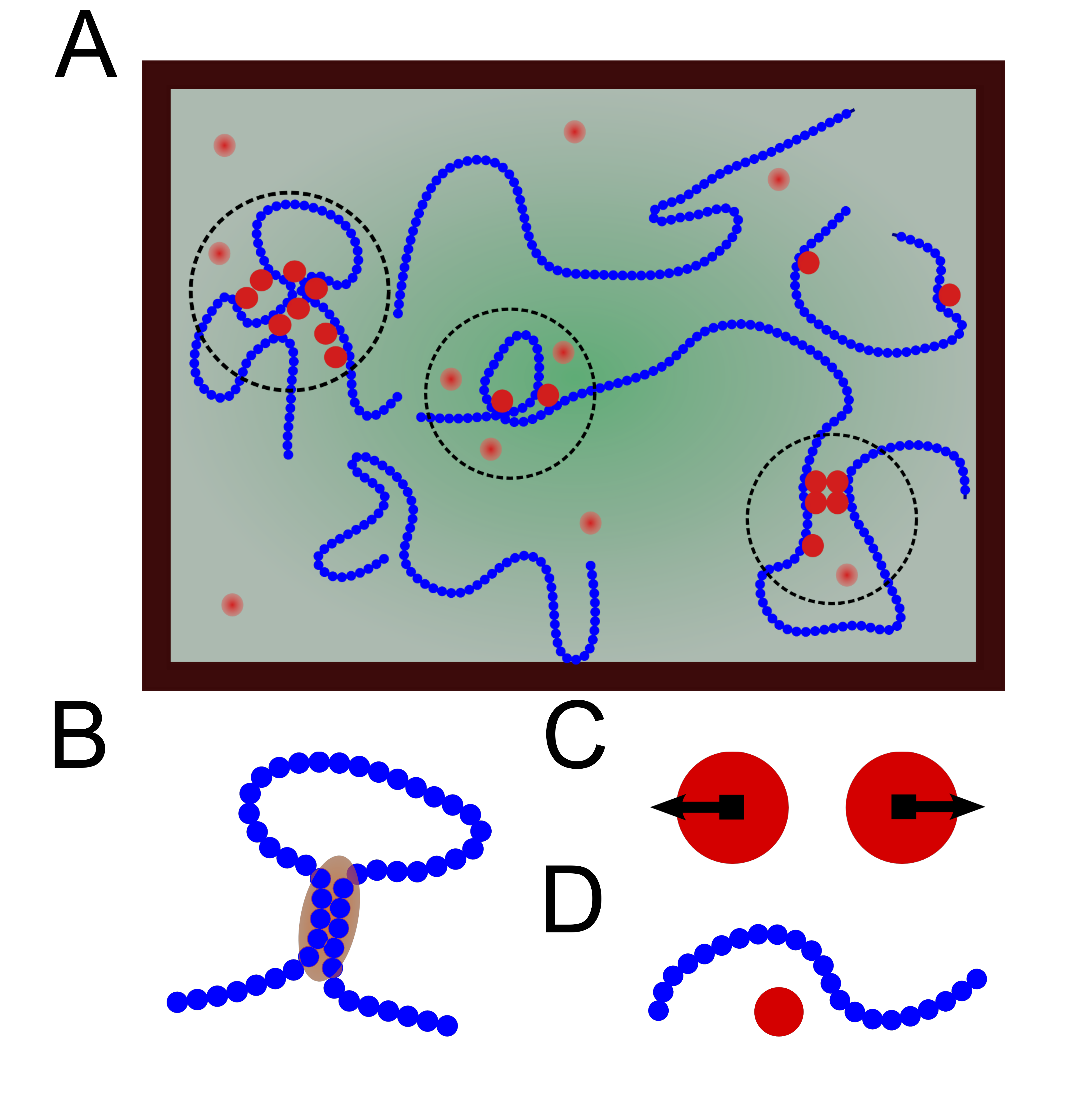}
\caption{\textbf{A}: model of DNA represented as beads-on-string polymers (blue) interacting with proteins (red). Dotted circles stand for clusters with high concentrations of DNA monomers and proteins. \textbf{B}: monomer-monomer interaction is repulsive. \textbf{C}: protein-protein interaction is repulsive. \textbf{D}: DNA-protein interaction is attractive.}
\label{fig::Figure1}
\end{figure}

Predicting the three-dimensional structure of chromosomes from the primary DNA sequence has become an important goal, as genomic and transcriptomic data are now generated at an elevated pace. In eukaryotes and prokaryotes, transcription of highly active genes has been shown through morphological evidence to occur within discrete foci containing RNA Polymerases (RNAPs).  It was later demonstrated that one given focal point was enriched in one type of dedicated transcription factor (TF) \cite{Llopis2010,Schoenfelder2010} and one type of gene promoter \cite{Cook2008}, as well as nascent transcripts \cite{Caudron-Herger2015}, thus justifying to name such foci "transcription factories". A thermodynamic model allowed to show that the stiff DNA polymer and properly located attractive sites mimicking TF bridges were necessary and sufficient ingredients to produce a transcription factory through DNA looping \cite{Junier2010}. Indeed, there is now convincing evidence that chromosomes are organized into loops (Hi-C, 3C, etc...) \cite{Schoenfelder2010}, and that looping brings distant genes together so that they can bind to elevated local concentrations of RNAPs (FISH, 3C, etc...) \cite{Spilianakis2005}. DNA-binding proteins such as TFs are generally positively charged, thus providing a non-specific interaction with the negatively charged DNA polymer. DNA sequence-dependent binding offers specific interactions. Together, non-specific and specific associations allow proteins to search their target DNA sequences more efficiently \textit{via} facilitated diffusion \cite{Sheinman2012,Berg1981a}, which combines three-dimensional diffusion in the bulk volume and mono-dimensional diffusion along the DNA. These considerations motivated studies to characterize the time-scale of the dynamics or anomalous diffusion. Molecular dynamics simulations are used to model proteins that diffuse to DNA, bind, and dissociate. The time scales reached in numerical simulations are usually several orders of magnitude smaller than the biological ones, and thus the phenomena observed during such simulations might be transient and irrelevant biologically. In this paper, we present a study of the properties and phase diagram of a DNA-protein solution, at thermodynamic equilibrium, which entails DNA condensation into compact structures induced by non-specific DNA-binding proteins. The calculated phase diagram is thus expected to be relevant at biological time scales. To do so, we will consider a simplified model in which the nucleus (or bacterial nucleoid) is represented by a closed volume $V$ (Fig. \ref{fig::Figure1}). The double-stranded DNA chains are modeled as $M$ semi-flexible polymer chains (polymerization index $N$) which interact with $P$ spheres, which represent either transcription factors or structural proteins. We consider the nucleus (or bacterial nucleoid) to be a good solvent for DNA chains, so that monomers experience a repulsive interaction between themselves. Conversely, we assume that there is an attractive interaction between proteins and DNA which allow the proteins to bind to DNA. As for the protein-protein interaction, we will consider a repulsive (hard-core) interaction, but the case of an attractive (\textit{e.g.} complexation, dimerization...) interaction could be treated in the same way. Finally, we make the assumption that all interactions are non-specific. In the sequel, subscripts $D$ and $P$ will stand for DNA and protein respectively. We will first describe the phase diagram of such a system in the mean-field approximation and show that there is a phase transition from a dilute phase at high temperature, to a concentrated phase of the DNA and proteins at lower temperature, which can be identified as the transcription factory phase. In a second step, we characterize the structure of the dense phase and show that it can adopt a crystalline order, suggesting an interesting parallel with the existence of some DNA biocrystals \textit{in vivo}. This method is general and can be applied to many genome architecture problems.

\section{Flory-Huggins theory}

\subsection{Free energy and thermodynamic functions}
In the following, we will study the phase diagram of the bulk of the bacterial cell (or nucleus) in the mean-field approximation. In the context of polymer theory, this approximation is also called the Flory-Huggins theory {\cite{deGennes1979}}. A similar kind of approach has been used to study the demixion of a mixture of polymers and colloids, in which the interaction is repulsive \cite{Sear2002}. By contrast, in the present work, the polymer-colloid interaction will be taken as attractive. We will denote by $c_D$ and $c_P$ the concentrations of DNA monomers and proteins, and by $\sigma_D$ and $\sigma_P$ the molecular volume of a DNA monomer and of a protein. The (Flory-Huggins) free energy per unit volume reads:
\begin{align} \label{eq::mean_field_free_energy_volumic}
\begin{split}
\beta f(c_D, c_P) &= \dfrac{1}{2} \alpha_D c_D^2 + \dfrac{1}{2} \alpha_P c_P^2 + v c_D c_P  + \dfrac{1}{6} w (c_D+c_P)^3 + c_P \log \dfrac{c_P \sigma_P}{e} + \dfrac{c_D}{N} \log \dfrac{c_D \sigma_D}{e N} \\
\end{split}
\end{align}
where $\alpha_D$, $\alpha_P$ and $v$ are second order virial coefficients denoting respectively the DNA-DNA, protein-protein and DNA-protein interactions, and $w$ is the third virial coefficient, necessary to avoid the collapse of the system. Note that this last term comes mostly from the entropy of the solvent. Indeed, if solvent molecules were present with concentration $c_S$ and molecular volume $\sigma_S$, the solvent translational entropy would be $c_S \log {c_S \sigma_S/e} $. If we assume incompressibility of the DNA-protein-solvent mixture (\textit{i.e.} $c_D + c_P + c_S = c_0$), the solvent entropy can be written in mean-field as $(c_0-c_D-c_P) \log {(c_0-c_D-c_P)\sigma_0/e}$ which, when expanded to 3rd order in $(c_D+c_P)$, yields the cubic term in Eq. {\ref{eq::mean_field_free_energy_volumic}}.

The Gibb's free energy per unit volume is the Legendre transform of Eq. \ref{eq::mean_field_free_energy_volumic}:
\begin{align} \label{eq::mean_field_gibbs_volumic}
\beta g(c_D, c_P)= \beta f(c_D, c_P) - \mu_D c_D - \mu_P c_P = -\beta \Pi
\end{align}
where $\Pi$ is the osmotic pressure, and where $\mu_D$ and $\mu_P$ are the chemical potentials of DNA monomers and proteins. The total number of particles of the system is fixed, but as we shall see later, these chemical potentials play a useful role when the system separates into two phases at equilibrium. At thermal equilibrium, the Gibb's energy is an extremum: $\partial \beta g / \partial c_D = 0$ and $\partial \beta g / \partial c_P = 0$, from which we deduce the chemical potentials:
\begin{align}
\begin{split}
\mu_D (c_D,c_P) &= \dfrac{\partial \beta f}{\partial c_D} = \alpha_D c_D + v c_P + \dfrac{1}{2} w (c_D + c_P)^2 + \dfrac{1}{N} \ln{\left( \dfrac{c_D \sigma_D}{N}\right) } \\
\mu_P (c_D,c_P) &= \dfrac{\partial \beta f}{\partial c_P} = \alpha_P c_P + v c_D + \dfrac{1}{2} w (c_D + c_P)^2 + \ln{(c_P \sigma_P)}
\end{split}
\end{align}

By inserting the last expressions in Eq. \ref{eq::mean_field_gibbs_volumic}, it is straightforward to see that $\alpha_D$, $\alpha_P$ and $v$ are indeed identified to the coefficients of a virial expansion.

The free energy in Eq. \ref{eq::mean_field_free_energy_volumic} is quite general and holds for arbitrary interactions between the constituents. It is possible to relate $\alpha_D$, $\alpha_P$ and $v$ to any pair potential, say $u(r)$, used to model the corresponding interaction. Indeed, these virial coefficients can be computed to lowest order (in the density) using the well known Mayer relation:
\begin{align} \label{eq::mayer_formula}
\alpha = - \int {\mathrm{d}^3 \mathbf{r} \, \left( e^{- \beta u(\mathbf{r})} - 1 \right)}
\end{align}
and the same for $v$. 

In physiological conditions, salt (\textit{e.g.} $NaCl$, $KCl$) and ions (\textit{e.g.} $Ca^{2+}$, $Mg^{2+}$) are present in solution, giving rise to screened electrostatic interactions. The interactions are therefore short-ranged with a range given by the Debye-H{\"u}ckel length. Yet , at the mean field level, the specific shapes of the interaction potentials is irrelevant, and the effect of ions in solution only arises through an adjustment of the Mayer coefficients $\alpha_D$, $\alpha_P$ and $v$. The DNA excluded volume coefficient $\alpha_D$, which accounts also for the electrostatic repulsion between negatively charged monomers is positive. The protein-protein coefficient $\alpha_P$ is in general positive (repulsive), due to electrostatic repulsion between identically charged proteins, but can be attractive (negative) for specific molecules, in particular they may undergo dimerization or hybridization. For hard spheres for instance, $\alpha = 2^3 \sigma$, $\sigma$ being the volume of one sphere. We hereafter restrict ourselves to the case where DNA-DNA and protein-protein interactions are purely repulsive (steric).

Conversely, we assume $v<0$, \textit{i.e.} the DNA-protein interaction has an attractive tail, which is temperature independent (in first approximation). As we shall see, this Flory-Huggins theory predicts the existence of a critical temperature $T^c$. We will assume that $v(T)$ is analytic in $\mid T - T^c \mid$ and can be written to leading order as:
\begin{align} \label{eq::flory_expression_v}
v(T) = v(T^c) \dfrac{\theta - T}{\theta - T^c}
\end{align}
where $\theta$ is the Flory temperature for which $v(\theta)=0$, i.e. the interaction vanishes. 

\subsection{The regime of phase separation}

As mentioned previously, there will be a phase transition when the homogeneous solutions $c_D=MN/V$ and $c_P=P/V$ become unstable. It is well known that generically, when the temperature decreases, the system separates into two phases. At the phase separation point, the homogeneous high temperature phase may stay metastable down to a point, called the spinodal point, where the homogeneous phase becomes totally unstable.  The so-called spinodal condition is given by the equation:
\begin{align} \label{eq::spinodal_condition}
\left| \dfrac{\partial^2(\beta f)}{\partial(c_D,c_P)} \right| = 
\left|
\begin{array}{l l}
\partial^2 \beta f / \partial c_D^2 & \partial^2 \beta f / \partial c_D \partial c_P \\
\partial^2 \beta f / \partial c_D \partial c_P & \partial^2 \beta f / \partial c_P^2 
\end{array}
\right| \le 0
\end{align}
where the array denotes the determinant of the matrix.

In general for $v(T)$ fixed, Eq. \ref{eq::spinodal_condition} with the equality determines a line of spinodal points, delimiting the region where the homogeneous mean field solution is stable from the region where it is not. 
In the unstable regions, the system undergoes a phase separation. 
If $T$ is increased, $v(T)$ becomes less negative. At some point, the spinodal lines merge into a point when $T$ reaches a critical value $T^c$ (Fig. \ref{fig::Figure2}). This is a tricritical point. For $T > T^c$ the homogeneous solution is stable for any value of $c_D^*$ and $c_P^*$, where we used the $\ast$ superscript to emphasize that these concentrations are the mean field solutions in the absence of phase separation. There are critical lines emerging from the tricritical point when the temperature is decreased, as will be seen later.

In a biphasic regime, the concentrations will be different but uniform in each of the two phases, separated by an interface whose energy is not extensive (the interfacial free energy is proportional to the surface of the interface). We label the dilute phase (resp. dense phase) by $I$ (resp. $II$). The total system free energy then reads:
\begin{align} \label{eq::mean_field_biphasic_free_energy}
\dfrac{\beta F^{tot}}{V} = \phi^{I} \beta f(I) + \phi^{II} \beta f(II)
\end{align}
where $f(I)$ is a short-hand for $f(c_D^{I}, c_P^{I})$ and $\phi^I$ and $\phi^{II}$ denote the volume fraction of the dilute and dense phase.

A straightforward minimization of Eq. \ref{eq::mean_field_biphasic_free_energy}, with the constraints of conservation of volume and particles number of $D$ and $P$, yields the usual equations of coexistence between phase $I$ and phase $II$:
\begin{align} \label{eq::mean_field_coexistence_equations}
\left\lbrace
\begin{array}{l l l}
\mu_D(I) &=& \mu_D(II) \\
\mu_P(I) &=& \mu_P(II) \\
\Pi(I) &=& \Pi(II) \\
\end{array}
\right.
\end{align}
where $\Pi$ denote the osmotic pressures of each phase.
The above equations are simply the equalities of the chemical potentials and the osmotic pressures. It trivially implies $\phi^{II} = 1 - \phi$ with $\phi = \phi^I$. Note that Eq. \ref{eq::mean_field_coexistence_equations} is a system of 3 equations with 5 variables $(c_P^I,c_D^I,c_P^{II},c_D^{II},T)$, thus it determines a surface of coexistence. 

\subsection{Results}

We present here the results of the mean field theory and defer the discussion of the actual values of the parameters. We will assume that DNA and protein spheres have the same size (diameter $a$) which we will use as the new unit length. The temperature $T$ and the coefficient $v(T)$ are related through Eq. \ref{eq::flory_expression_v}, and we therefore introduce the order parameter $t$:
\begin{equation} \label{eq::order_parameter}
v(T) = v(T^c) \left( 1 + t \right)
\end{equation}

\begin{figure}[ht!]
\centering
\includegraphics[width= 1 \linewidth]{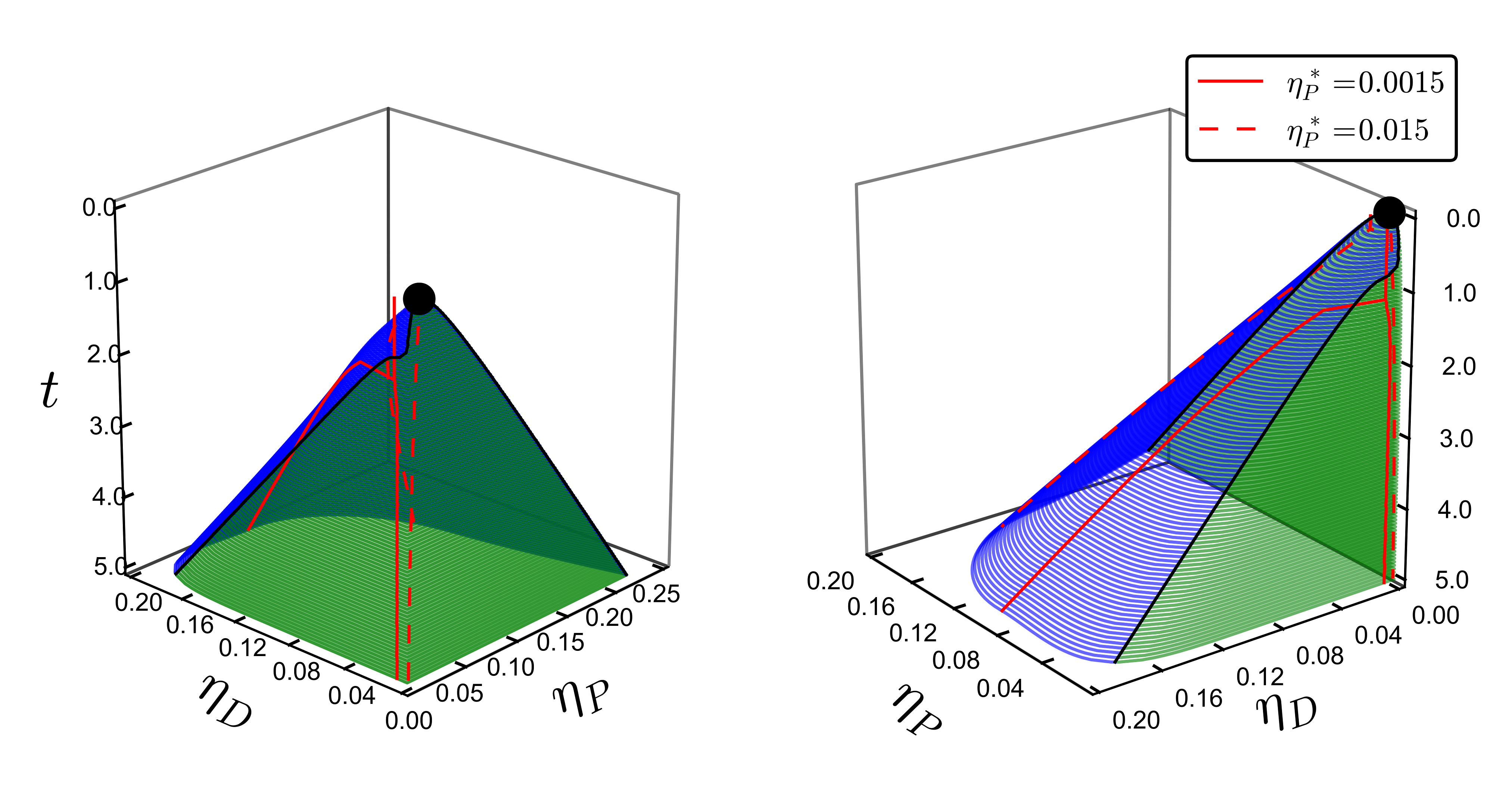}
\caption{3D representation of the coexistence surface in a $ \left(\eta_D, \, \eta_P, \, t \right)$ coordinate system.  Coexistence lines are shown for $\eta_P^*=0.0015$ and $0.015$, with $\eta_D^*= 0.01$ (red lines). Critical lines (black lines) emerge from the tricritical point (black dot).}
\label{fig::Figure2}
\end{figure}

We choose to discuss the phase separation in terms of $t$ and of the densities $\eta_D =  c_D \sigma_D$ and $\eta_P = c_P \sigma_P$. The coexistence surface is then computed numerically by solving Eq. \ref{eq::mean_field_coexistence_equations} and is shown in Fig. \ref{fig::Figure2}. At the critical point, Eq. \ref{eq::mean_field_coexistence_equations} has a unique solution, namely the triplet $(\eta_D^c, \eta_P^c, T^c)$. As previously reported for such systems \cite{Sear2002}, we find that
\begin{align}
\eta_D^c \sim \dfrac{1}{\sqrt{N}}
\end{align}

This is the same scaling as the overlap density for polymer chains of size $N$, which is the density at which polymer coils begin to interpenetrate (see for instance \cite{deGennes1979}). Given that DNA is present in the nucleoid at concentrations close to the overlap density, this suggests that biological systems may function at the vicinity of this tricritical point.

\begin{figure}[ht!]
\centering
\includegraphics[width= 0.48 \linewidth]{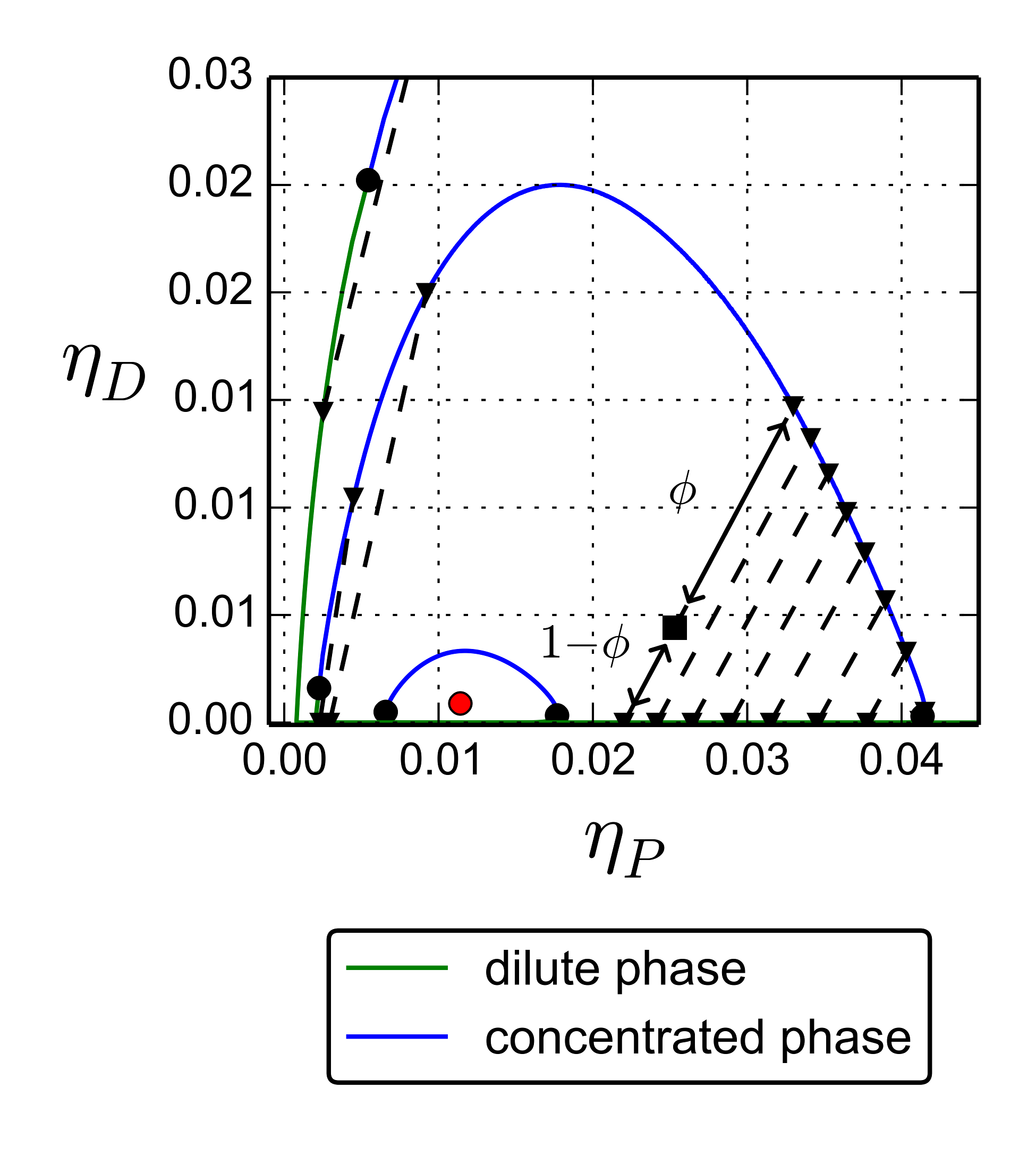}
\caption{Coexistence lines for $t = 0.05, \, 0.5, \, 1.0$. The coexistence line shrinks toward the tricritical point (red dot) when $t \to 0$. For each curve, the dilute phase is shown in green and the concentrated phase is shown in blue. Coexisting states are connected by tie lines (dotted segments). The volume fraction of each phase is determined (black arrows) according to Eq. \ref{eq::mean_field_maxwell_rule}.}
\label{fig::Figure3}
\end{figure}

For $T < T^c$ we have $ v(T)  < v(T^c) < 0$ which corresponds to a DNA-protein interaction more attractive than at the critical point and the solutions of Eq. \ref{eq::mean_field_coexistence_equations} are distributed on a closed curve. This closed curve is in fact the collection of all pairs of coexisting dilute $(c_D^I,c_P^I)$ and dense $(c_D^{II},c_P^{II})$ phases. They are represented in Fig. \ref{fig::Figure3}, where the pairs of coexisting phases are connected by tie lines. Although there is an infinite set of possible states of coexistence, the total number of DNA and protein spheres selects a unique solution pair. The resulting coexistence state is uniquely determined by the following relation, whose graphical interpretation is shown in Fig. \ref{fig::Figure3}:
\begin{align} \label{eq::mean_field_maxwell_rule}
\phi \left(\begin{array}{l} c_D^{I} \\ c_P^{I} \end{array} \right) + (1 - \phi) \left(\begin{array}{l} c_D^{II} \\ c_P^{II} \end{array} \right) = \left(\begin{array}{l} c_D^{*} = MN / V \\ c_P^{*} = P / V \end{array} \right)
\end{align}

An important point to note is that in general, the concentration of DNA in the dilute phase is very small.
Indeed, the translational entropy of DNA is small, due to the factor $1/N$, and thus there is no entropic gain for the DNA to be in the dilute phase, whereas it has an important enthalpic advantage to be in the concentrated phase. To illustrate how one can determine the composition of the system, we now explain how phase separation takes place when cooling the system from high to low temperature. We consider the case of DNA and protein densities:
\begin{align} \label{eq::physiological_densities}
\eta_D^*	&=0.01\\
\eta_P^*	&=0.0015
\end{align}

\begin{figure}[ht!]
\centering
\includegraphics[width= 1 \linewidth]{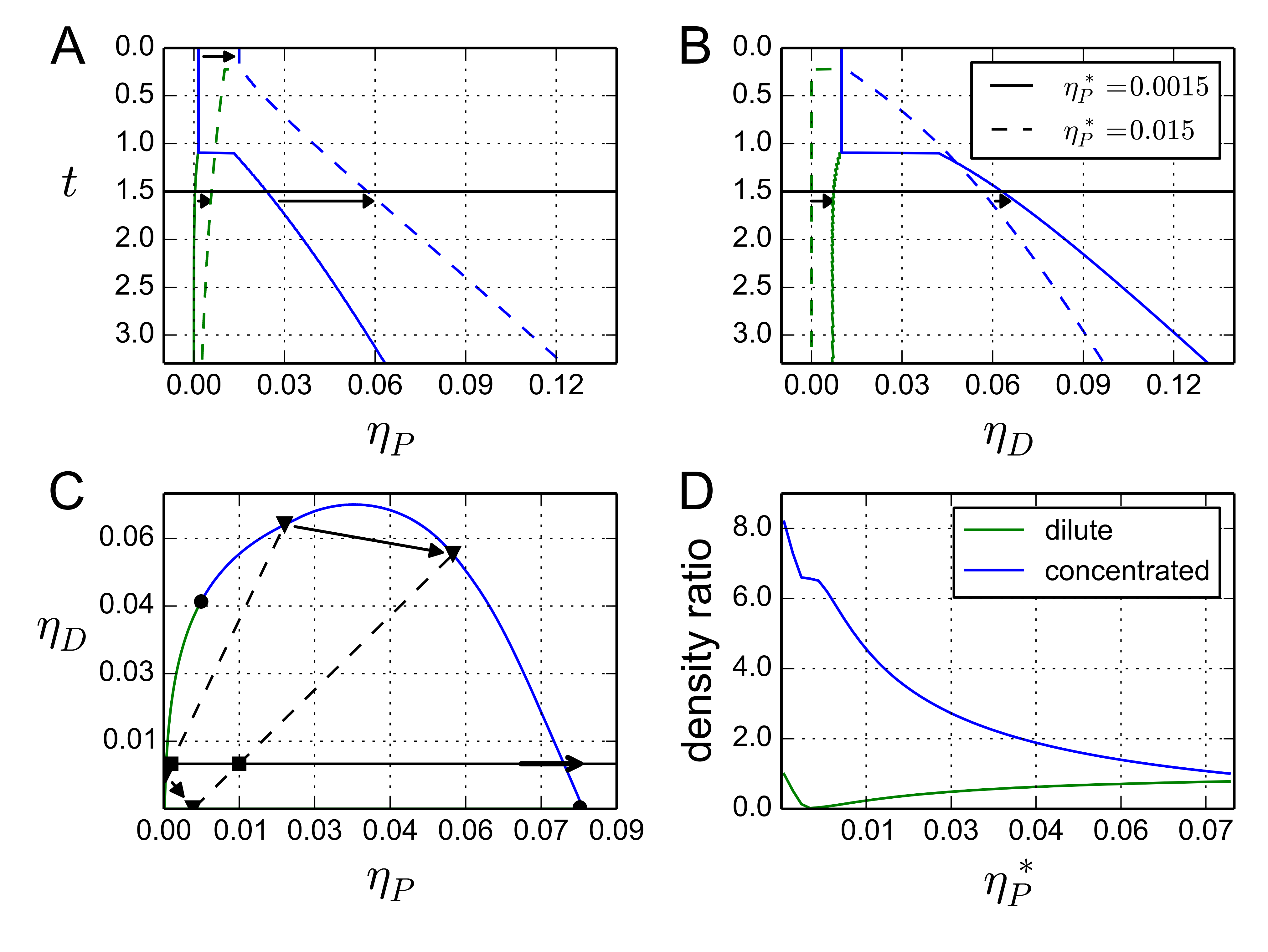}
\caption{\textbf{A} and \textbf{B}: coexistence lines for $\eta_P^*=0.0015$ and $0.015$, with $\eta_D^*= 0.01$. These are the projections of the coexistence lines of Fig. \ref{fig::Figure2} as a function of the density of proteins (\textbf{A}) or DNA (\textbf{B}). \textbf{C}: section of the phase diagram at $t=1.5$ (corresponding to the parameters given in Tab. \ref{tab::lennard_jones_parameters}). A path corresponding to a fixed DNA density of $\eta_D^*= 0.01$ is drawn (black line). \textbf{D}: density ratio $(\eta_D + \eta_P) / (\eta_D^* + \eta_P^*)$ for the dilute phase and dense phase, for a density of DNA fixed to  $\eta_D^*= 0.01$.}
\label{fig::Figure4}
\end{figure}

Again, we defer the justification for this choice of the parameters. The system splits into two phases at a temperature $T^*$. For $T>T^*$ the system is homogeneous with  values of the concentration given by Eq. \ref{eq::physiological_densities}, whereas for $T<T^*$, the system splits into two phases whose composition is determined by Eq. \ref{eq::mean_field_coexistence_equations} and \ref{eq::mean_field_maxwell_rule}. The line of coexistence obtained is shown on Fig. \ref{fig::Figure2}. Note that at $T=T^*$, the phase transition is first order, except when $T^*$ is on a critical line, in which case it is second order. Let us now assume that the concentration of proteins is increased by a factor of ten to $\eta_P^*=1.5 \, 10^{-2}$. When the system is cooled from high temperatures, it splits into two phases, and as before, the coexisting states are distributed on the surface of coexistence. However the mass conservation requirement (Eq. \ref{eq::mean_field_maxwell_rule}) yields a different line of coexistence. The new line of coexistence is shown with dashed lines in Fig. \ref{fig::Figure2} and Fig. \ref{fig::Figure4}. As might have been expected, an augmentation of the protein concentration results in an increased protein concentration in both the dilute and concentrated phase (Fig. \ref{fig::Figure4}A). The DNA concentration, however, shows a two-step pattern. When further proteins are added to the solution, the "free" DNA monomers of the dilute phase are transferred to the concentrated one, and consequently, the DNA concentration in the dense phase first increases. But at some point, the DNA concentration in the dense phase reaches a maximum and starts to decrease (Fig. \ref{fig::Figure4}B-C). Indeed, a fraction of the newly added proteins will populate the concentrated phase and make it swell, while the amount of DNA remains the same. Therefore varying the total quantity of proteins can induce non monotonous variations of the DNA concentration in the phases of the system.

\section{Structure of the dense phase}
\subsection{Example of structures}

\begin{figure}[ht!]
\centering
\includegraphics[width= 0.48 \linewidth]{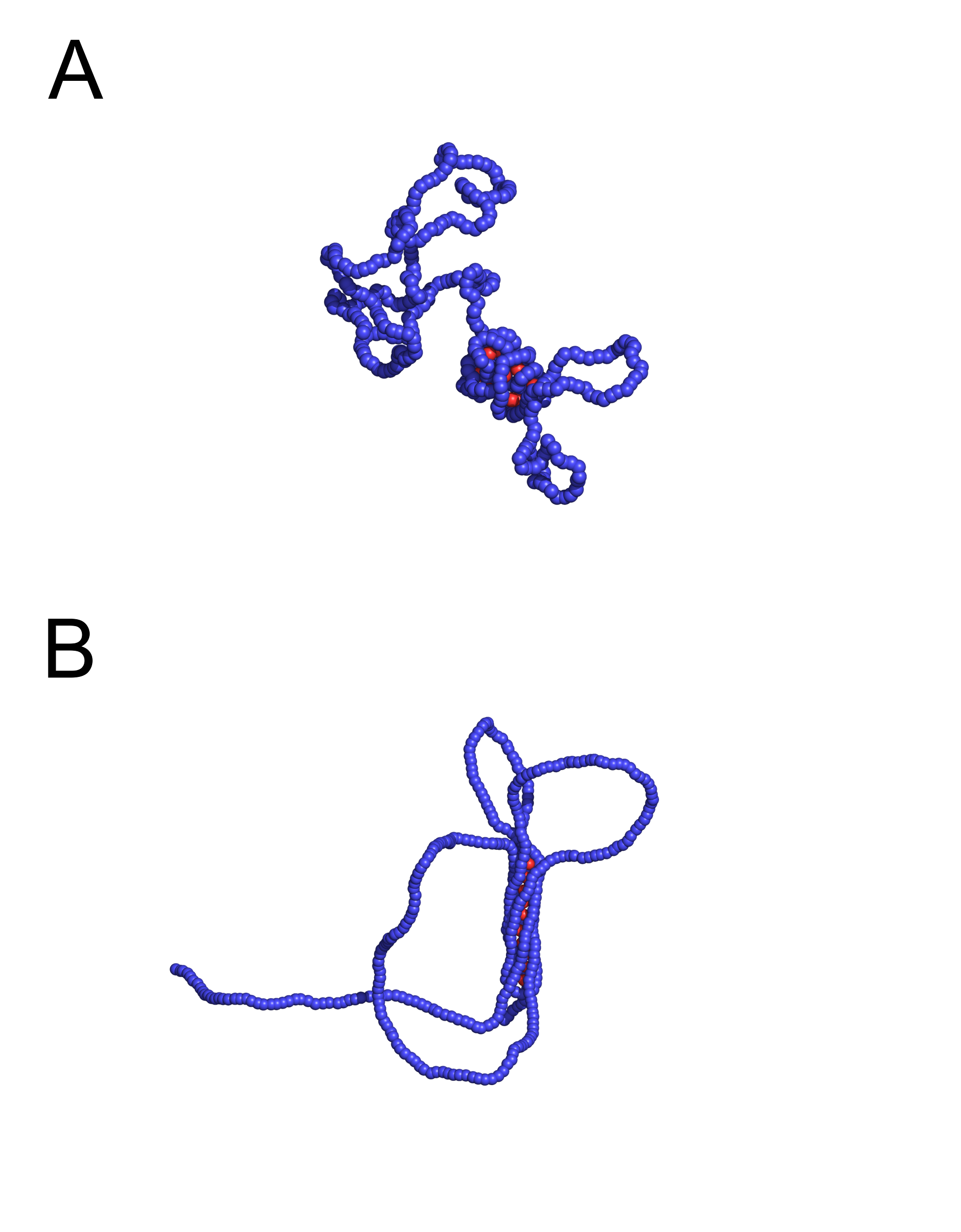}
\caption{Two equilibrium configurations of a single polymer chain (blue) displaying the coexistence of a dense and a dilute phase, with persistence length $l_p=1$ (\textbf{A}),  and $l_p=20$ (\textbf{B}) interacting with proteins (red). For small bending rigidity, the structure of the dense phase is globular whereas it is cylindrical in the other case. We performed simulations with $P=10$ spheres (see methods).}
\label{fig::Figure5}
\end{figure}

In the last section, we saw that the Flory-Huggins theory predicts the existence of a phase separation between two homogeneous phases. It is well known that within the Flory-Huggins approximation, the chain structure is not taken into account, except through the suppression of the translational entropy of the chains. In particular, the fact that chains may have a strong bending rigidity (long persistence length) does not play any role at this level. Therefore, the predicted structure of the dense phase is that of a melt of collapsed polymer with spheres. However, several studies have highlighted that the bending rigidity of the polymer has an influence on the microstructure of the dense phase \cite{Brackley:2013aa,Marenduzzo2015,Orland1996}. This is well characterized (Fig. \ref{fig::Figure5}). A standard way to characterize the effect of the chain structure is to use the Random Phase Approximation (RPA) (see reference \cite{deGennes1979}). We have performed such RPA calculations, but we don't report it here, because it did not show any interesting instability in the dense phase. The reason for the failure of RPA is that the phase transition from the homogeneous to the separated phases is first order and thus is not driven by critical fluctuations. This is a typical case when RPA fails to give insights about the dense phase structure.

\subsection{Theory of Hamiltonian paths}
\begin{figure}[ht!]
\centering
\includegraphics[width= 0.48 \linewidth]{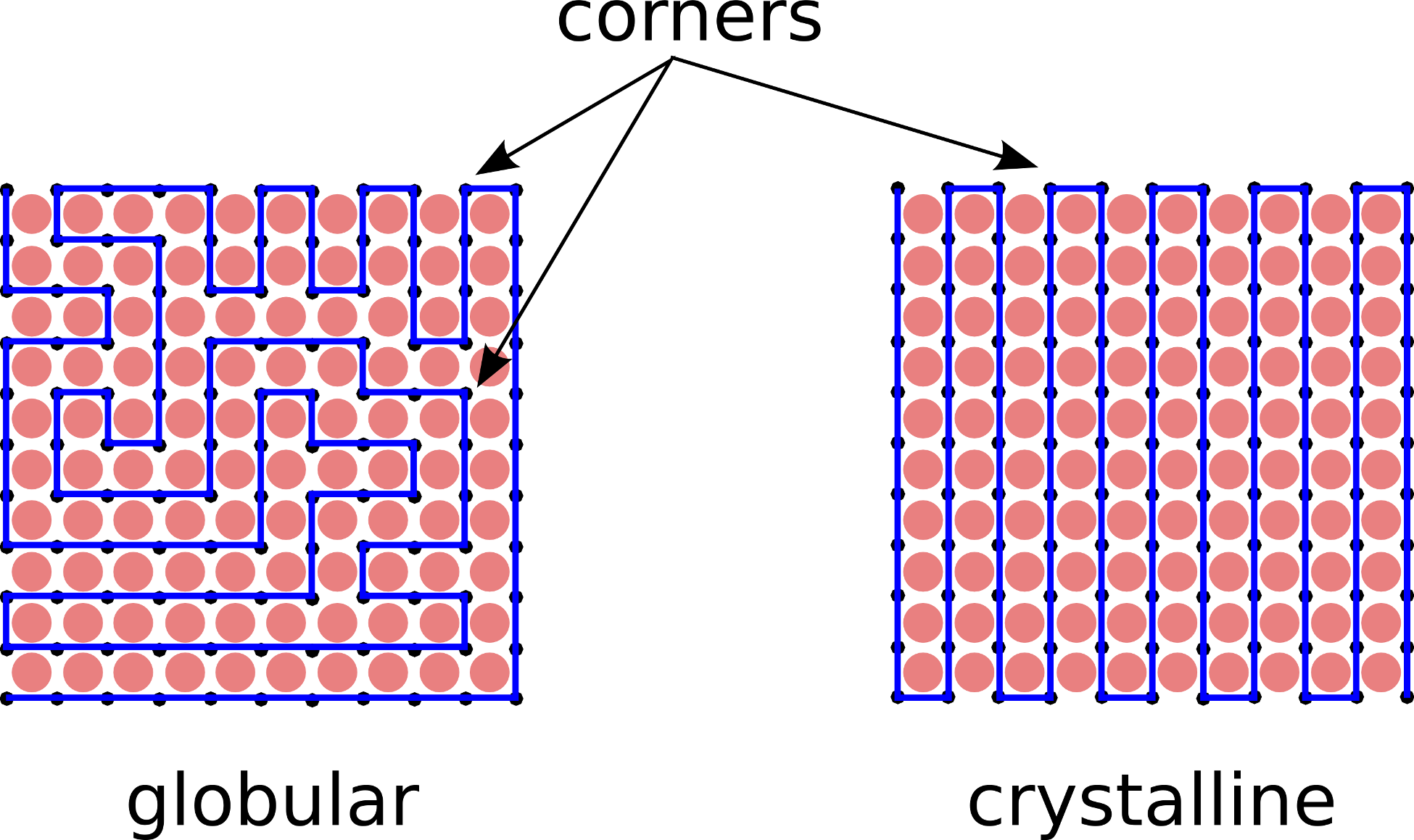}
\caption{Two realization of Hamiltonian paths on a cubic lattice. The globular state contains an extensive number of corners whereas the crystalline state contains a non-extensive number of corners (proportional to the surface).}
\label{fig::Figure6}
\end{figure}

Since RPA is not appropriate to describe the system in the dense phase, we adopt an other approach. Because of their attractive interactions with the DNA, the spheres in which the polymer is immersed, play the role of colloid particles which bridge various parts of the polymers. Consequently, these spheres induce an effective attraction between the monomers. 
We thus turn to a model of a semi-flexible polymer chain on a lattice that has been proposed initially to explain the folding of a protein in compact structures \cite{Orland1996,Orland1992,Orland1993} (see Fig. \ref{fig::Figure6}). An attraction energy $\epsilon_v$ between non-bonded nearest neighbors is included, which favors compact configurations. A bending energy of the chain is introduced as a corner penalty. It penalizes corners by an energy $\varepsilon_h$ and thus plays the role of a bending rigidity. As we will see, this term induces an ordering transition between a random (molten) globule where corners are mobile in the bulk, and a crystalline phase, where corners are expelled to the surface of the globule. Using a mean field theory, it was shown that depending on the temperature and chain stiffness, three phases can exist, namely a dilute phase where the polymer is swollen, a condensed phase, which we call a molten globule, where the polymer is collapsed and disordered  and finally a second condensed phase where the polymer is collapsed but with a local crystalline ordering. The phase diagram is described simply by the two parameters $\epsilon_v$ and $\epsilon_h$. For fixed small $\epsilon_h$, there is a second-order phase transition at a temperature $T=T_\theta$ between a dilute and a disordered condensed phase, followed by a first-order freezing  transition at $T_F$ between the disordered condensed phase and a locally ordered condensed phase of the polymer. Upon increasing the chain stiffness $\epsilon_h$, the molten globule region shrinks until it eventually vanishes. For larger stiffness, the polymer goes abruptly from a swollen to a frozen configuration ($T_F > T_\theta$) through a direct first order transition (Fig. \ref{fig::Figure7}). These theoretical results were readily confirmed and improved by Monte-Carlo simulations \cite{Orland1996,Grassberger2008,Frenkel1998}.

In its simplest form, this model considers a completely collapsed polymer on a lattice. It is represented as a Hamiltonian path (HP) on a lattice, that is a path which visits each site once and only once. Thus the density of monomer is $\eta=1$. A good approximation to the total number ${\cal N}$ of HP on a lattice was shown to be \cite{Orland1985}:
\begin{equation} \label{eq::hamiltonian_path}
{\cal N} = \left(\frac{q}{e}\right)^N
\end{equation}
where $N$ is the total number of points of the lattice, and $q$ is the coordination number of the lattice, \textit{e.g.} $q=2d$ on a $d$-dimensional cubic lattice.

In the case of semi-flexible polymers (HP with corner penalty), the partition function is \cite{Orland1996}:
\begin{align}
\mathcal{Z} = \sum \limits_{ \{ HP \} } e^{-\beta \epsilon_h {N}_C(HP)}
\label{eq::HP_partition_function}
\end{align}
where ${N}_C(HP)$ counts the number of corners of a HP realization.

A saddle-point approximation gives the corresponding free energy per monomer:
\begin{align} \label{eq::free_energy_HP}
\beta f = - \ln \dfrac{q(\beta)}{e}
\end{align}
where
\begin{align}
q(\beta)=2 + 2(d-1)e^{-\beta \epsilon_h} 
\end{align}
is an effective coordination number,
$e=2.718\, 28 ...$ and $d$ is the dimensionality of the lattice. Note that if the corner penalty vanishes, we recover the result of Eq. \ref{eq::hamiltonian_path}. 

As the temperature decreases, the effective coordination number $q(\beta)$ decreases, and the free energy increases. There is a temperature $T_F$ for which $q(\beta)=e$ giving a free energy per monomer $f(T_F)=0$. For $T<T_F$, $f(T)$ would become positive in Eq. \ref{eq::free_energy_HP} if the saddle-point approximation were still valid. However $f(T$) is a negative quantity \cite{Orland1992} and therefore remains zero below the freezing temperature $T_F$. Consequently for temperatures $T>T_F$, corners are mobile in the bulk, leading to a liquid-like structure for the corners, whereas for $T<T_F$, the polymer is frozen in stretched configurations with $f(T)=0$, in which corners are expelled to the surface and polymer segments tend to be aligned inside (Fig. \ref{fig::Figure6}). These configurations have been studied previously: they are elongated neck structures or toro{\"i}ds \cite{MacKintosh2004}, whose typical size is given by
\begin{align}
\dfrac{\epsilon_h}{U(\beta)} \sim l_p
\end{align}
where $U(\beta) = \partial (\beta f) / \partial \beta$ is the internal energy.

This simple model can be extended to the case where the volume fraction $\eta < 1$. It requires the introduction of the parameter $\epsilon_v = -\chi$, where $\chi$ is the Flory parameter of a polymer chain in a solvent and denotes the effective attraction between monomers induced by the proteins. This leads to the computation of the above mentioned phase diagram in terms of the two parameters $\epsilon_v$ and $\epsilon_h$.

\subsection{Results}
To make the Hamiltonian paths approach more quantitative, it is interesting to relate its parameters to our simplified picture of DNA interacting with proteins. Namely, we would like to relate $\epsilon_v$ and $\epsilon_h$ to the parameters of the Flory-Huggins free energy in Eq. \ref{eq::mean_field_free_energy_volumic}. But in the last one, the monomer-monomer attraction is mediated by spheres. In the dense phase, the total concentration of monomers and spheres: $c=c_D + c_P$ is essentially fixed to the close packing concentration $c_0$. By inserting this in Eq. \ref{eq::mean_field_free_energy_volumic} we obtain
\begin{align}
\beta f = \left(\dfrac{\alpha_D + \alpha_P}{2} + v(T) \right) c_D^2 + \dfrac{1}{6} c^3 + (c-c_D) \ln\dfrac{c-c_D}{e}
\end{align}
where we neglected the translational entropy of the polymer in the dense phase and dropped the linear terms in $c_D$ as this results in an adjustment of the chemical potentials. We then have the correspondence:
\begin{align}
\epsilon_v \equiv - v(T) - \dfrac{\alpha_D + \alpha_P}{2}
\end{align}

For low temperature, $v(T)$ can reach large negative values. This is mapped to a large $\epsilon_v$ in the HP model. Depending on the rigidity of the chain, the dense phase might be globular (low $l_p$) or crystalline (large $l_p$). 
The effective monomer density in the dense phase is given by
\begin{align}
\eta \equiv \dfrac{c_D}{c_D + c_P}
\end{align}

One important result obtained using a HP model is the phase diagram of a polymer on a lattice (implicit solvent) with bending rigidity, obtained in mean-field in ref. \cite{Orland1996} and then supplemented by Monte Carlo studies \cite{Grassberger2008}. We show here that the phase diagram of a semi-flexible polymer interacting explicitely with spheres in an off-lattice volume is very similar. We performed Brownian Dynamics simulations with a polymer chain of $N=400$ beads and $P=100$ protein spheres in a cubic volume of size $L=100$ with periodic boundary conditions (see methods). Polymer beads and protein spheres interact through a Lennard-Jones potential with a well depth given by the energy scale $\epsilon$ (in $k_B T$). We used a Kratky-Porod model of polymer, with bending rigidity characterized by the persistence length $l_p$. By varying $l_p$ and $\epsilon$ independently, we were able to explore the phase behaviour of this system. We monitored the coil-globule transition by looking at the quantity:
\begin{align}
q=\dfrac{\log R_g}{\log N}
\end{align}
where $R_g$ is the radius of gyration of the polymer. For a self-avoiding polymer with scaling law $R_g \sim b N^{\nu}$, $q = \nu + cst / \log N$. In a good solvent, the polymer is swollen with $\nu=0.588$ whereas in a bad solvent it collapses with $\nu=1/3$ . It is clear that $q$ varies like $\nu$.

Following the same authors \cite{Grassberger2008}, we define the quantity $n_\alpha = \sum \mid \mathbf{u}_i \cdot \mathbf{e}_\alpha \mid$ for $\alpha=x,y,z$, in which $i$ runs over all the bonds of the polymer, $\mathbf{u}_i$ is the unit vector having the same direction as the bond $i$ and $\mathbf{e}_\alpha$ is the unit vector of the corresponding $\alpha$-axis. We then define $n_{min}=\min_{\alpha} (n_\alpha)$, $n_{max}=\max_{\alpha} (n_\alpha)$ and
\begin{align}
p=1 - \dfrac{n_{min}}{n_{max}}
\end{align}
For an isotropic configuration, $n_x=n_y=n_z$ resulting in $p=0$. Conversely, for a configuration streched in one direction, say along the x-axis, $n_x=1$ and $n_y=n_z=0$, resulting in $p=1$. Thus $p$ measures the directional order of the polymer.

\begin{figure}[ht!]
\centering
\includegraphics[width= 0.75 \linewidth]{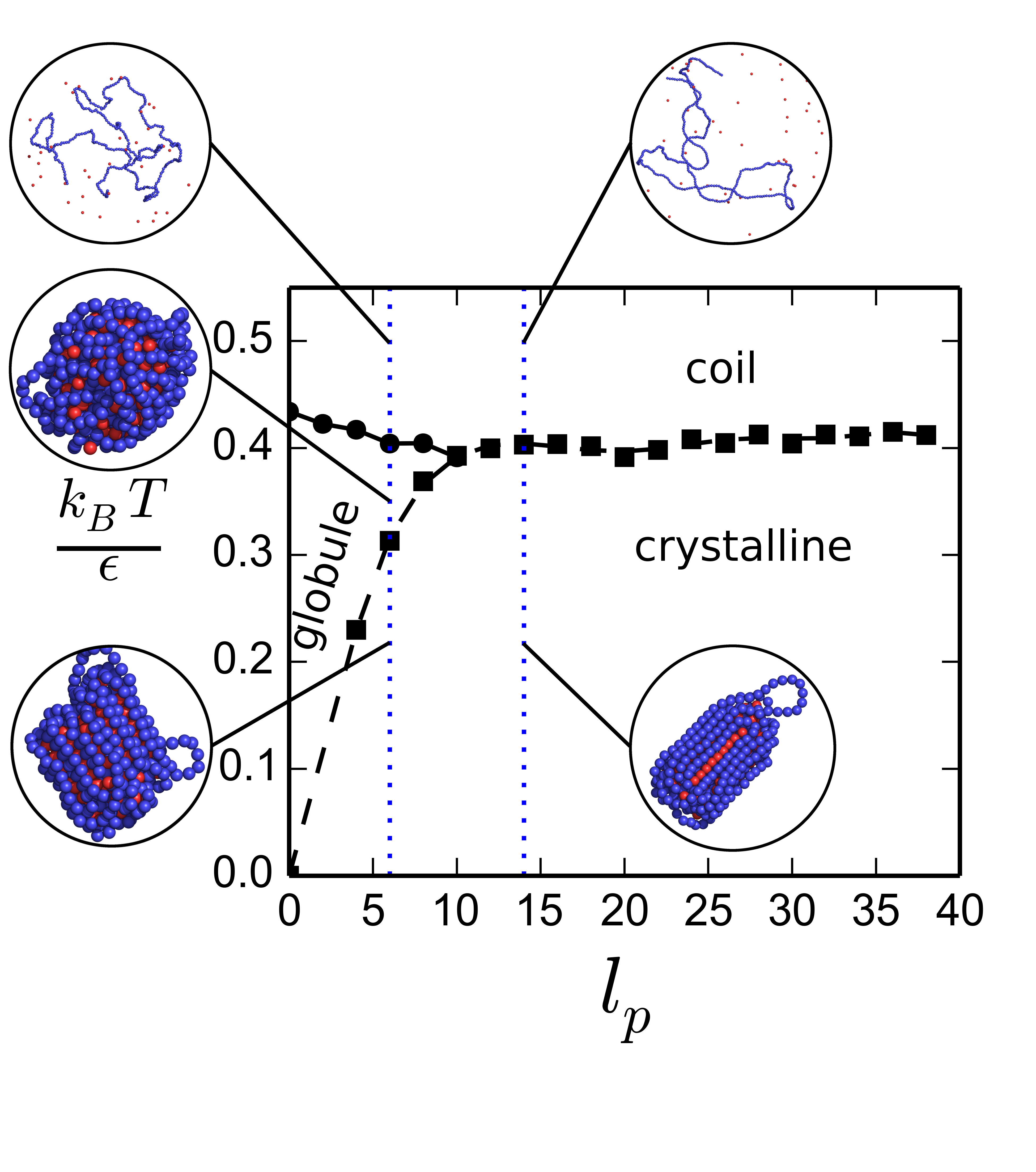}
\caption{Phase diagram obtained for a polymer chain interacting with spheres. The phase diagram is plotted as a function of $k_B T /\epsilon$ and $l_p$, where $\epsilon$ is the strength of the Lennard-Jones DNA-protein interation and $l_p$ is the persistence length. We performed simulations with $P=100$ spheres (see methods).}
\label{fig::Figure7}
\end{figure}

We plotted the phase diagram obtained as a function of $k_B T / \epsilon$ and $l_p$ (Fig. \ref{fig::Figure7}). There is a clear similarity with the case of a polymer on a lattice without explicit proteins. However we observe that the coil-globule and globule-crystal transitions occur at a higher interaction energy $\epsilon$. This might be a consequence of going from a lattice model to a continuous model. It might also be due to the fact that in the high temperature regime, the concentration of spheres in solution is smaller than the close packing, therefore making it hardly comparable to an actual solvent. There is a specific persistence length $l_p^c \simeq 10$ such that:
\begin{itemize}
\item	for $l_p < l_p^c$, the polymer collapses through a second order coil-globule transition, followed by a first order globule-crystal transition when $\epsilon$ increases;
\item	for $l_p > l_p^c$, the coil-globule transition no longer exists and the polymer collapses directly from a coil to a crystalline phase through a first order phase transition.
\end{itemize}

 \section{Parameters and methods}
\paragraph{Molecular dynamics simulations\newline}

In order to sample configurations of our system we used the LAMMPS (Large-Scale Atomic / Molecular Massively Parallel Simulator) software package. The system is coarse-grained so that DNA is modeled as a beads-on-string polymer and proteins as spheres. The simulations are run in the Brownian Dynamics (BD) mode. The resulting dynamics consists in the integration of the Langevin equation:
\begin{align}
m_i \dfrac{\mathrm{d}^2 \mathbf{r}_i}{\mathrm{d}t^2} = \mathbf{F}_i - \gamma_i \dfrac{\mathrm{d} \mathbf{r}_i}{\mathrm{d}t} + \sqrt{2 k_B T \gamma_i} \mathbf{\eta}_i (t)
\end{align}
where $i$ is the index of one bead, $\mathbf{r}_i$ is the position of the bead, $\gamma_i$ the friction coefficient, $\mathbf{F}_i$ is the resulting force exerted on the bead from the rest of the system, $k_B$ is the Boltzmann constant and $T$ is the temperature. The last term is a stochastic force in which $\mathbf{\eta}_i(t)$ is a Gaussian white noise such as: $\left\langle \mathbf{\eta}_i(t) \mathbf{\eta}_j(t') \right\rangle=\delta_{ij} \delta(t-t')$. Simulations were run with $m_i=1$, $\gamma_i=1$, $k_B T=1$, in a cubic volume of size $L=100$ with periodic boundary conditions and we took a polymer with $N=400$ beads. The diameter $a$ of the beads was taken as the unit length.

As for the polymer chain model, beads $i$ and $i+1$ are connected with a finetely extensible non-linear potential (FENE) such that:
\begin{align}
U_{el} (\mathbf{u}_i) = -\dfrac{K_{el}}{2}  r_0^2 \ln{\left( 1 - \dfrac{u_i^2}{r_0^2}  \right)}
\end{align}
where $\mathbf{u}_i=\mathbf{r}_i - \mathbf{r}_{i-1}$. Simulations were run with $K_{el}= 30 \, k_B T / a^2$ and $r_0=1.5 \, a$.

The bending rigidity of the polymer chain is introduced through the Kratky-Porod potential:
\begin{align}
U_{b}(\theta_i)= K_{b} \left(1 - \cos \theta_i \right)
\end{align}
where $\theta_i$ is the angle between vectors $\mathbf{u}_i$ and $\mathbf{u}_{i+1}$. The bending coefficient is related to the persistence length by $K_b=l_p \times k_B T$.

We restricted our analysis to the case of DNA monomers and protein spheres of same dimension (diameter $a$). This assumption does not alter the main qualitative features of the physics but makes the discussion simpler. Steric as well as attractive interactions between beads are introduced using a truncated Lennard-Jones potential such as:
\begin{align}
U_{LJ}(\mathbf{r}_{ij}) = 
\left\lbrace
\begin{array}{l l}
4 \epsilon \left[ \left(\dfrac{a}{r_{ij}} \right)^{12} -\left(\dfrac{a}{r_{ij}} \right)^6 - \left(\dfrac{a}{d^{tr}} \right)^{12} +\left(\dfrac{a}{d^{tr}} \right)^6 \right] & \text{ if } r_{ij} < d^{tr} \\
0 & \text{ otherwise }
\end{array} \right.
\end{align}

\begin{figure}[ht!]
\centering
\includegraphics[width= 0.48 \linewidth]{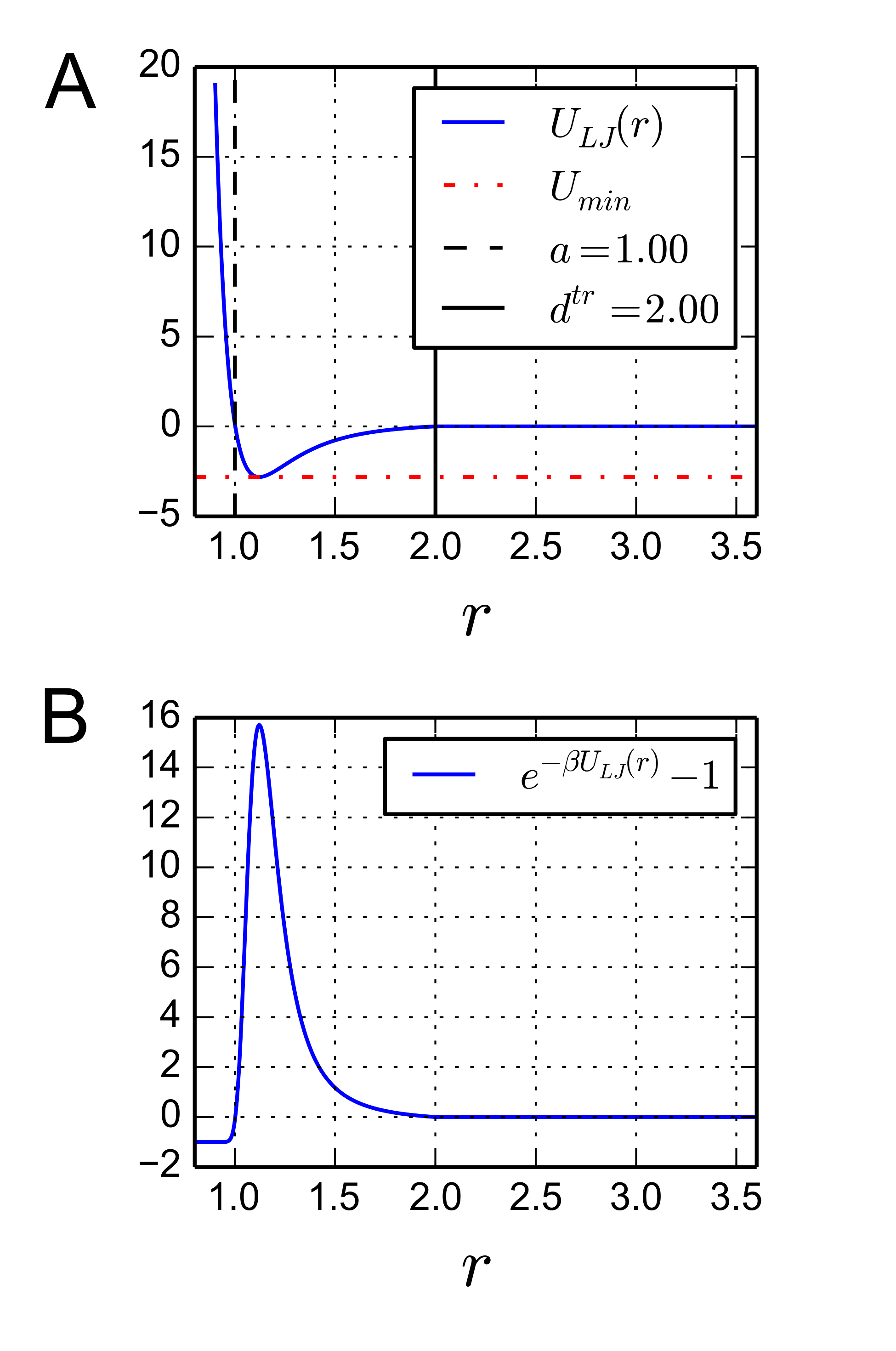}
\caption{Truncated Lennard-Jones potential modeling an attractive interaction between DNA and proteins with parameters indicated in Tab. \ref{tab::lennard_jones_parameters}. \textbf{A}: potential shape. \textbf{B}: Mayer function.}
\label{fig::Figure8}
\end{figure}

Unless stated otherwise, we took the values indicated in Tab. \ref{tab::lennard_jones_parameters}. The DNA-DNA interaction being purely repulsive, we retrieve $\alpha_D \simeq 2^3 \sigma_D$ where $\sigma_D$ is the volume of one monomer. The same argument holds for the proteins. Conversely, the monomer-sphere coefficient $v(T)$ is largely negative (Tab. \ref{tab::lennard_jones_parameters} and Fig. \ref{fig::Figure8}).

\begin{table}[ht]
\begin{tabular}{| l l | c c c| c |}
\hline
1	& 	2 	&	$\epsilon$	&	$a$		&	$d^{tr}$	&	Mayer coefficient	$\alpha$ \\
\hline
DNA	&	DNA	&	$1.00$		&	$1.00$		&	$1.12$			&	$4.40$			\\

Protein	&	Protein	&	$1.00$		&	$1.00$		&	$1.12$			&	$4.40$	\\

DNA	&	Protein	&	$3.00$		&	$1.00$		&	$2.00$			&	$-62.6$	\\
\hline
\end{tabular}
\caption{Parameters for the truncated Lennard-Jones potential modeling the DNA-DNA, protein-protein and DNA-protein interactions. The mean field pair coefficients are computed using Mayer formula.}
\label{tab::lennard_jones_parameters}
\end{table}

We systematically performed equilibration runs of $10^5$ time units to reach thermal equilibrium. We then performed sampling runs of $10^6$ time units which we used to measure physical quantities. One simulation time units can be understood as follows. The diffusion coefficient can be expressed as $D=a^2 /\tau_B$ where $\tau_B$ is the Brownian time corresponding to one simulation time unit. Therefore, we can do a mapping to physical units. Assuming a diffusion coefficient of $D \simeq 10 \, \mu m^2 s^{-1}$  for a protein in the bacterial nucleoid \cite{Elowitz1999}, we find $\tau_B = 600 \, ns$ for $a=6 \, nm$ and $\tau_B=2.0 \, \mu s$ for $a=20 \, nm$.

\paragraph{Mean-field concentrations\newline}
To assess physiological concentrations, we assimilate an \textit{E. coli} bacteria to a cylinder of radius $0.5 \, \mu m$ and length $1 \, \mu m$. We took a genome length of $4.6 \, 10^6$ base pairs ($bp$). We consider that the diameter of one monomer is $1 \, a = 6 \, nm \equiv 17 \, bp $, so that one such chromosome is modeled as a polymer of $N=2.6 \, 10^5$ monomers of diameter $1 \, a$. This leads to a density of DNA $\eta_D \sim 10^{-2}$. As for the proteins, there are several DNA-binding proteins, called nucleoid-associated proteins, which act on the structure of the DNA. In \textit{E. coli} for instance HU, H-NS, Fis, RecA, Dps and other proteins have this stucturing function. We choose to use Fis as a reference because it is a well known structural protein with a large number of target sites \cite{Kahramanoglou2011} and binds to DNA sequences of $17$ nucleotides \cite{Nowak-Lovato2013}. In \textit{E. coli}, there are approximately $75,000$ Fis proteins per genome copy \cite{Ishihama2014} in the early exponential growth phase, yielding a ratio $\eta_P / \eta_D \simeq 0.3$. This is also consistent with previous numerical studies for which $0.1 < \eta_P / \eta_D < 0.5$ \cite{Marenduzzo2015}. Eventually we suggest that this ratio might be quite general and can also be mapped to eukaryotes. Indeed, we could either consider a human cell (genome length $\sim 3.3 \, 10^9 \, bp$ and nucleus diameter of $10 \, \mu m$) and take $1 \, a = 20 \, nm$ as a unit length (size of protein complex). Considering that the DNA would be packed as euchromatin with linear packing fraction $\nu = 100 \, bp/nm$ \cite{Brackley:2013aa}, the unit length is $1 \, a \equiv 1.7 10^6 \, bp$ and  we get $\eta_D \sim 10^{-2}$. In the former case the proteins are assumed to be roughly $6 \, nm$ in diameter while the latter case better describes the interaction of DNA with large protein complexes.

\begin{figure}[ht!]
\centering
\includegraphics[width=0.48 \linewidth]{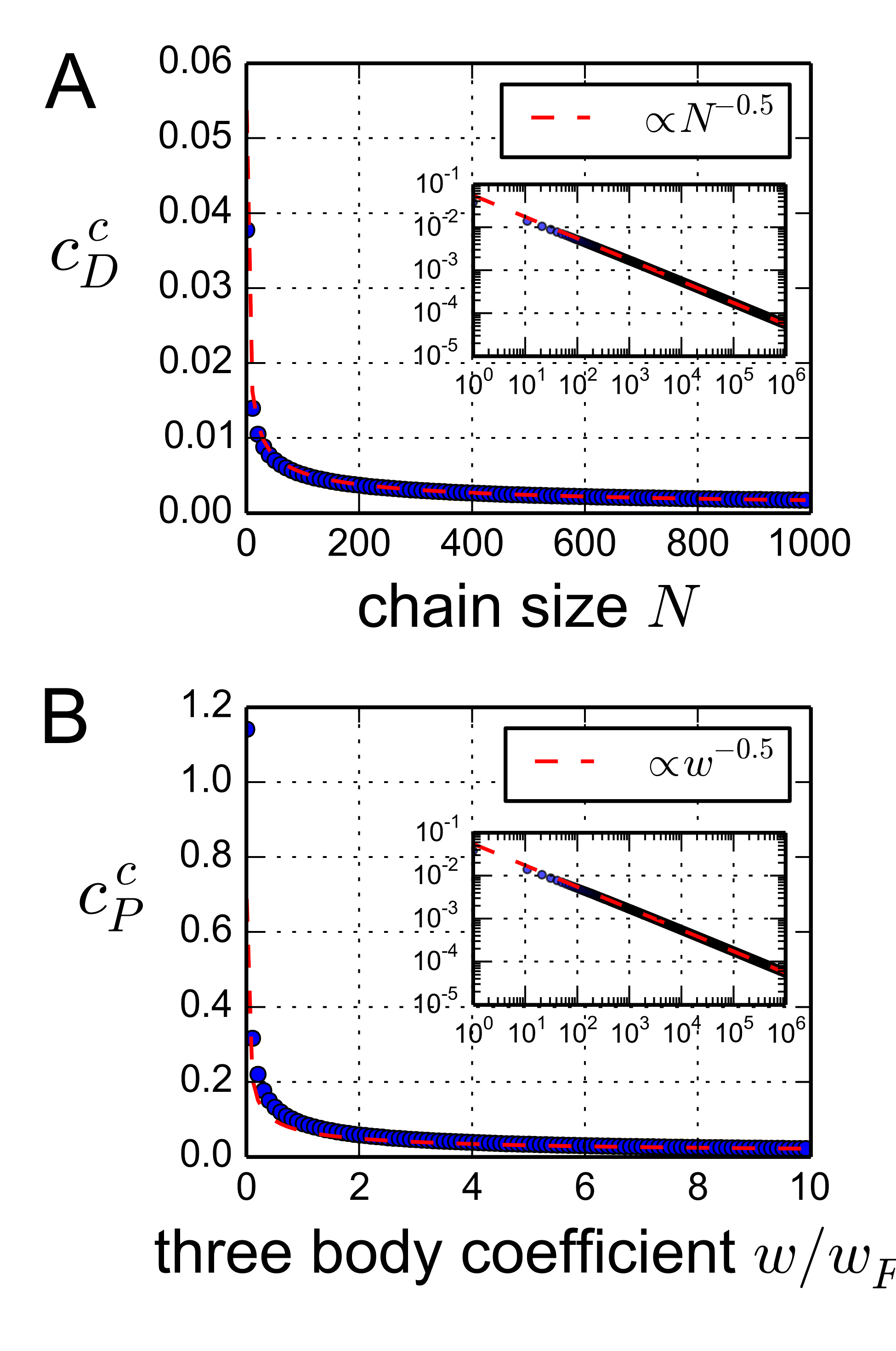}
\caption{\textbf{A}: the DNA critical concentration $c_D^c \sim 1 / \sqrt{N}$. \textbf{B}: for $w > 10 w_F$ the variations of $c_P^c$ and of the other critical parameters are very slow.}
\label{fig::Figure9}
\end{figure}

\paragraph{Computation of the phase diagram \newline}
The coexistence surface is computed by solving Eq. \ref{eq::mean_field_coexistence_equations}. Due to numerical accuracy limitations, we computed the phase diagrams for chains with polymerization index $N=5000$. Since the critical values scale like $1 / \sqrt{N}$ (Fig. \ref{fig::Figure9}), this arbitrary choice captures the essential features of the $N\to \infty$ limit. Finally, the model relies on a free parameter $w$, which can be used to fit the model. In Flory theory, $w$ is extracted from the development of the entropy of mixing and would be $w_F = {1}/{c_0^2}$, where $c_0$ is the close packing concentration. However cells are crowded environments, containing other species, metabolites or organelles. Furthermore, biological compartments can also restrict the accessible volume. Eventually, this value is underestimated and we arbitrarily choose $w =  10 \times w_F$. Although it could be much larger, the variations of $c_P^c$ above this value are quite slow and will not alter dramatically the previous phase diagrams (Fig. \ref{fig::Figure9}). With all these parameters, we find for the tricritical point coordinates: $c_P^c = 2.2 \, 10^{-2} \, a^{-3}$, $c_D^c = 7.7 \, 10^{-4} \, a^{-3}$ and $\beta^c v(T^c)=-25.67 \, a^{3}$.

\section{Discussion and Conclusion}

\paragraph{DNA condensation \textit{in vitro}\newline}

The condensation of DNA induced by DNA-binding proteins or ions has been thoroughly studied. It is well known that DNA collapses from disperse structures corresponding to swollen coil configurations into ordered, highly condensed states. This has been the focus of several \textit{in vitro} experimental works \cite{Bloomfield1996,Livolant1996,Vasilevskaya1995,Yoshikawa1995,Durand1992}. One important conclusion from these studies is that during its collapse, DNA undergoes phase transitions through the following three phases: isotropic fluid, cholesteric and crystalline (hexagonal), in agreement with our results. As stated above, within the Flory-Huggins theory, the phase transition induced by ions or DNA-binding proteins appears to be first order, except at the tricritical point and on the critical lines. Therefore the transition from the swollen to the condensed state should be discontinuous and present hysterisis effects, which was indeed observed \cite{Yoshikawa1995,Widom1980}. Interestingly, the Flory-Huggins theory predicts another effect. At fixed temperature,  there is a line of possible coexisting states. Given a certain amount of DNA let us discuss the consequences of adding proteins to the system. The protein concentration would increase from zero until it reaches a value for which the system splits into two phases. If we keep adding proteins, the system will at some point exit the biphasic regime (Fig. \ref{fig::Figure4}C-D). This phenomenon called reentrance has been observed in some experimental work \cite{Vasilevskaya1995}.

\paragraph{DNA condensation \textit{in vivo}\newline}
Although it is premature to draw any clear biological conclusion, it is tempting to discuss at least qualitatively the effect of DNA condensation on biological functions. In eukaryotes, nucleosomal organization provides an effective protection against detrimental factors. This organization is absent in prokaryotes, which have a significantly lower ratio of DNA-binding proteins \cite{Kellenberger1992}. However, in harsh environmental conditions (radiations, temperature, oxidating agents and radicals), several bacteria resort to DNA condensation mechanisms to protect their genome. Maybe the most spectacular case is the appearance of macroscopical DNA aggregates with crystal-like order in starved \textit{E. coli} cells. In stressful conditions, the alternative $\sigma^S$ factor is expressed, in response to low temperature, cell surface stress or oxidative shock. This in turn induces the expression of the DNA-binding protein Dps \cite{Almiron1992,Almiron1994}. In starved cells, Dps is the most abundant DNA-binding protein, with approximately 20,000 Dps protein per cell. Consequently, DNA is condensed into crystal-like aggregates, which make it less accessible to damaging factors. This process is reversible and wild-type \textit{E. coli} cells starved for three days remain unaffected by a high dose of oxidating agents whereas mutants lacking Dps lose viability \cite{Almiron1992}. Interestingly, Dps binds non-specifically to DNA. In regard to what has been discussed in this paper, we may infer that when Dps concentration increases, a dense phase appears. But at a scale of Dps size ($\sim 10 \, nm$), the apparent rigidity of DNA is large ($\sim 50 \, nm$). Therefore, as seen in Fig. \ref{fig::Figure7}, we might be in a case where the coil-globule transition is precluded by the freezing transition. Other examples of DNA compaction by non-specific proteins seem to exist \cite{FrenkielKrispin2006,Newton1996}.

\paragraph{Local concentration effects and transcription\newline}

Increasing evidence suggests that transcription proceeds from nucleation points called transcription factories, which are formed from the interaction of DNA with general and dedicated transcription factors. Although the non-specific hypothesis is not guaranteed, it is true that RNAP can bind widely onto DNA thanks to its $\sigma$-unit. The Flory-Huggins results from this paper suggest that a biphasic regime can exist, with a dense phase spanning a volume of size $(1-\phi) V$ and with local concentrations of DNA and RNAP increased by a factor of 4-to-8 with respect to the mean-field ones (Fig. \ref{fig::Figure4}D). This would result in shifting the equilibrium of complexation reactions such as:
$$
\text{ DNA } + \text{ protein } \rightleftarrows RNAP \text{ bound to } DNA
$$
towards the formation of complexes and may favour transcription initiation. This is consistent with some experimental work showing that RNAP clusters are formed during preinitiation and initiation of transcription \cite{Darzacq2013}. The same authors also proposed that crowding of enzymes, \textit{i.e.} higher local concentrations, may aid in rate-limiting steps of gene regulation. From a dynamical standpoint, the confinment of unbound RNAP in a restricted volume of size $(1-\phi)V$ can reduce the search time for a promoter. To this extent, it is worthwhile to point out a recent study claiming that the promoter search mechanism is indeed dominated by 3D diffusion of RNAP over the 1D diffusion along DNA \cite{Wang2013}.

\paragraph{Structure of the dense phase\newline}
Earlier studies have demonstrated that the frozen phase can present various metastable states \cite{Frenkel1998}. In the $N \to \infty$ limit ($N$ is the length of one chain), the transition time scale from one to another could be very large, and the system might well never equilibrate within biological timescales. Finally, the parallel drawn between the Hamiltonian paths theory and the Flory-Huggins theory does not pretend to mathematical rigor. One essential difference is that in our case the attractive interaction between monomers is mediated by spheres. A way to compute more precisely the structure of the dense phase would be to go beyond the homogeneous saddle point approximation, for instance by using the so-called  self-consistent field theory method \cite{Edwards1988,Fredrickson2005}, which is a very complex method in the case of semi-flexible polymers.

\paragraph{Conclusion\newline}

We presented here two complementary frameworks to describe the phase diagram of polymeric fluids induced by colloids, and applied it to a DNA chain interacting with DNA-binding proteins. Starting from a Flory-Huggins free energy, we first computed the mean-field phase diagram and found that at low temperature (\textit{i.e.} high DNA-protein affinity) a biphasic regime exists, consisting of the coexistence of a dilute phase and a concentrated phase. The dilute phase may correspond to swollen configurations of the DNA whereas the concentrated phase is a model for condensed states of DNA. The theory may also apply to DNA condensation by multivalent ions or proteins in general. Second, we addressed the characterization of the dense phase structure and showed that the chain bending rigidity can have dramatic effects. Without bending rigidity, the dense phase has no directional order and is a molten globule. However, when the chain bending rigidity is large enough, there is a freezing transition from the globular to crytalline phase. Eventually for very rigid chains, the coil-globule transition is precluded by the freezing transition and the phase transition predicted in the Flory-Huggins framework does not occur.

In the cell, the existence of a dense phase could be a good approximation for the transcription factories observed experimentally. It is conjectured that this may increase the rate of success in transcription initiation  by means of protein crowding and by enhancing the promoter search mechanism. Note that at a scale coarse-grained to several thousand base-pairs (gene scale), the chromosome is flexible and the dense phase has the structure of a molten globule. Conversely, at a scale of a few base-pairs, the apparent rigidity of DNA is much higher. Thus, the Dps protein, which binds non-specifically to DNA, can induce the collapse of the \textit{E. coli} chromosome into crystal-like aggregates; the dense phase is then frozen. This is not an efficient state for a searching mechanism. But on the contrary, it is very adequate to protect DNA.

The two frameworks are quite general and can be used to describe biological phenomena where DNA compaction occurs under the cooperative effect of binding proteins. In the future, we plan to apply it to other biological cases when more quantitative experiments become available.

\section{Acknowledgments}
The authors thank the MEGA team members at iSSB for excellent discussions. This work was supported by the IDEX Paris-Saclay grant, CNRS Genopole and the ANR project "Synpathic".

\footnotesize
\bibliographystyle{letreut} 
\bibliography{paper_biblio}

\end{document}